# Tuning the Interfacial Charge, Orbital and Spin Polarization Properties in $La_{0.67}Sr_{0.33}MnO_3$/ $La_{1-x}Sr_xMnO_3$ Bilayers

*Santiago J. Carreira[1,2,3,§], Myriam H. Aguirre[4,5,6,†], Javier Briatico[7,‡], Eugen Weschke[8,◊], and Laura B. Steren[1,2,3,\*]*

[1]Insituto de Nanociencia y Nanotecnología, B1650 San Martín, Buenos Aires, Argentina.

[2]Consejo Nacional de Investigaciones Científicas y Técnicas, C1425FQB Ciudad Autónoma de Buenos Aires, Argentina.

[3]Dpto. Materia Condensada, Centro Atómico Constituyentes, B1650 San Martín, Buenos Aires, Argentina.

[4]Instituto de Nanociencia de Aragón, Universidad de Zaragoza, E-50018 Zaragoza, Spain.

[5]Departamento de Física de la Materia Condensada, Universidad de Zaragoza, E-50009 Zaragoza, Spain.

[6]Laboratorio de Microscopías Avanzadas, Universidad de Zaragoza, E-50018 Zaragoza, Spain.

[7]Unité Mixte de Physique, CNRS, Thales, Université Paris-Sud, Université Paris-Saclay, Palaiseau 91767, France.

[8]Helmholtz-Zentrum Berlin für Materialienund Energie, Wilhelm-Conrad-Röntgen-Campus BESSY II, Albert-Einstein-Strasse 15, D-12489 Berlin, Germany.

[§]sjcarreira@gmail.com

[†]maguirre@unizar.es

[‡]javier.briatico@thalesgroup.com

[◊]eugen.weschke@helmholtz-berlin.de

*steren@tandar.gov.ar (corresponding author).

The possibility of controlling the interfacial properties of artificial oxide heterostructures is still attracting researchers in the field of materials engineering. Here, we used surface sensitive techniques and high-resolution transmission electron microscopy to investigate the evolution of the surface spin-polarization and lattice strains across the interfaces between $La_{0.66}Sr_{0.33}MnO_3$ thin films and low-doped manganites as capping layers. We have been able to finely tune the interfacial spin-polarization by changing the capping layer thickness and composition. The spin-polarization was found to be highest at a critical capping thickness that depends on the Sr doping. We explain the non-trivial magnetic profile by the combined effect of two mechanisms. On one hand, the extra carriers supplied by the low-doped manganites that tend to compensate the overdoped interface, favouring locally a ferromagnetic double-exchange coupling. On the other hand, the evolution from a tensile-strained structure of the inner layers to a compressed structure at the surface that changes gradually the orbital occupation and hybridization of the 3d-Mn orbitals, being detrimental for the spin polarization. The finding of an intrinsic spin-polarization at the A-site cation observed in XMCD measurements reveals also the existence of a complex magnetic configuration at the interface, different from the magnetic phases observed at the inner layers.

# 1. Introduction

Significant progresses made in the field of interface engineering along the last years were possible due to technological advances in the fabrication of thin films and multilayers and important developments in the field of surface sensitive techniques.[1-3] The possibility of studying and controlling oxide interfaces is still intriguing researchers owing to their potential applications in spintronic devices among other technologies.[4,5] Transition metal-oxide compounds are considered a model of strongly-correlated electron materials, in which the charge, orbital and spin degrees of freedom are extremely sensitive to strain and surface symmetry breaking effects.[6-8] The $La_{0.67}Sr_{0.33}MnO_3$ (LSMO) ferromagnet has been usually chosen to integrate into magnetic tunnel junctions as electrode, exploiting its highly spin-polarized magnetism mediated by a double-exchange mechanism. Nevertheless, the tunneling magnetoresistance (TMR) ratios obtained in junctions made with this manganite are much smaller than expected.[9] The low TMR values were attributed to a depression of the magnetism near the interfaces between electrodes and the insulating barrier.[10] This behavior could be related to a modified electronic structure originated at surface-symmetry breaking effects.[11,12] In an attempt of avoiding the formation of a dead layer at the electrode/barrier interface several authors proposed the use of low Sr-doped manganites $La_{1-x}Sr_xMnO_3$ (x<0.1) (LSxMO) as barrier layers.[10-13] The ground state of the LSxMO bulk compounds exhibits an insulating character and antiferromagnetic or weak ferromagnetic order at low temperatures.[14] However, in thin films the undoped member of this family $LaMnO_3$ (LMO) was reported to show robust ferromagnetism.[15] The interface of LSMO/LMO bilayers attracted special attention due to its higher spin-polarization with respect to that of LSMO/STO ones.[13,16] In contrast to the STO barrier layer, LMO may act as an electron donating layer stabilizing the ferromagnetic phase at the interface.[16] The coexistence of different magnetic phases as embedded clusters or the presence of distinct magnetic layers at oxide interfaces are today a matter of discussion, showing that there are still open issues to resolve.[17-19] Within this frame and with the goal of providing a deeper insight into the magnetism, charge and orbital reconstructions of oxide interfaces we investigated LSMO/LSxMO (*t*) (x: 0, 0.1; 3u.c. ≤ *t* ≤ 15u.c.) bilayers using high-resolution transmission electron microscopy and x-ray absorption and magnetic dichroic spectroscopies.

## 2. Experimental Results

2.1 Crystalline structure and strain analysis

The STEM-HAADF images of STO//LSMO/LSxMO reveal an epitaxial and coherent growth without misfit dislocations (**Figure 1**). In the STEM-HAADF image shown in Figure 1(a) the manganite films are clearly differentiated from the STO substrate. On the contrary, the contrast between the barrier and the electrode is hardly visible due to the close similarity in the composition of both layers. However, the barrier appears slightly brighter than the electrode in all the images due to its higher La content. The contrast in the composition was further confirmed with an EELS scan along the c-axis (see Figure S1 and S2 in the Supp. Info. for more details).

In order to quantify the strain state at the structure, the Geometrical Phase Analysis (GPA) was applied to the HRSTEM-HAADF images.[20] The in-plane strain-map, calculated along the *a*-axis, parallel to the interface, (Figure 1(b)), shows that there are distortions homogeneously distributed and no differences between substrate and films are appreciated. Instead, the strain map obtained for the direction perpendicular to the samples' surface shows that the LSMO electrode is contracted around 1% with respect to the $SrTiO_3$ due to tensile strains induced by the substrate while the LSxMO barrier is enlarged around 2%. The structural disruption at the LSMO/LSxMO interface is originated at the differences in the La/Sr ratio content between the barrier and the electrode that leads to a transition from a tensile (c<a) to compressive (c>a) strained crystalline structure along the perpendicular to the surface direction.

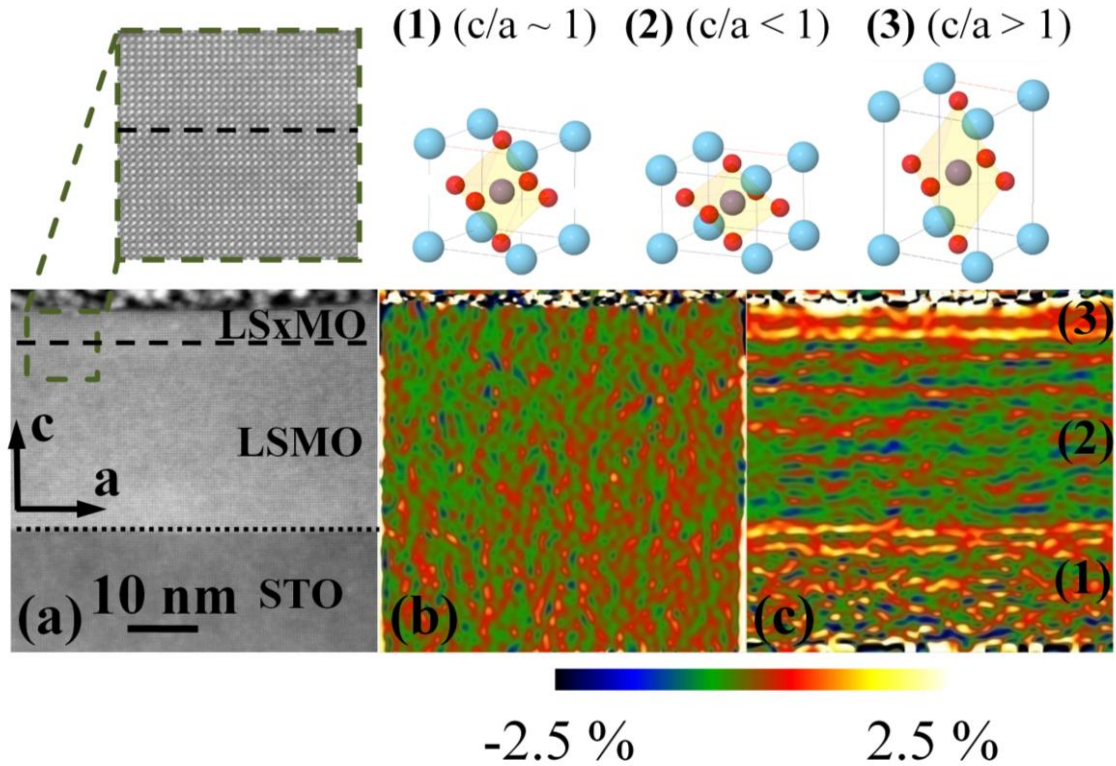

**Figure 1.** (a) Cross sectional STEM-HAADF image of a LSxMO/LSMO//STO bilayer. The STO/LSMO interface is denoted with a dotted line whereas the LSMO/LSxMO interface is indicated with a dashed line. HRSTEM-HAADF-GPA strain analysis maps of the components (b) parallel and (c) perpendicular to the interfaces.

2.2 Charge and magnetic reconstruction at the interface.

The XAS spectra at the Mn $L_{2,3}$- edge absorption presents two broad and well separated peaks $L_3$ and $L_2$, corresponding to $2p_{3/2} \rightarrow 3d_{5/2}$ and $2p_{1/2} \rightarrow 3d_{3/2}$ transitions, respectively (Figure S3 on the Supp. Info.).[21] Both the energy peak shifts and the peak intensity ratio $L_3/L_2$ are the fingerprints of the valence balance of Mn at the outer layers. In Figure 2 we plotted the peak position $Mn^{3+/4+}$, taken from the XAS spectra as a function of the barrier thickness and for different temperatures. The Mn valence state tends to decrease as the barrier thickness increases as a consequence of the larger contribution of the low-doped or undoped manganite to the signal. The influence of the Sr-doping on the barrier becomes evident by comparing Figure 2(a) and 2(b), i.e. a more sharp variation of the $L_3$ position is observed for the t-LMO samples with respect to the ones measured for t-LS0.1MO ones. This behavior is in agreement with the compositional profile obtained with the EELS scans (Figure S1 and S2) and can be understood by a different charge accommodation in the two series of samples that depends on the contrast of $Mn^{3+}/Mn^{4+}$ content at the electrode/barrier interfaces.

The effects of the spacer thickness on the interfacial ferromagnetic moment are clearly revealed by the XMCD spectra. The excellent signal/noise ratio of the measurements and the similarities between the XMCD lineshapes (see Figure S4 on the Supp. Info.) allowed us to make a comparative study of the magnetization in terms of the barrier thickness. According to this analysis,[22] the integral over the XMCD spectra was used to calculate the manganese moments in the two set of samples (Figure 2c-d). To make this analysis, we took into account that the escape depth of the electrons ($\lambda \sim 4$ nm) and the barrier thicknesses (1.2 nm $< t <$ 6 nm) are of the same order of magnitude.[23] Our results show that the magnetic moments vary with temperature, barrier thickness and Sr-doping. At 4 K, the magnetic moments increases with the barrier thickness, reaching a maximum value at a critical thickness $t_M$ of around 2.7 nm and 3.5 nm for the t-LMO series and t-LS0.1MO series, respectively. These results put in evidence that a few layers of low-doped manganites over the LSMO electrodes induces important enhancements of the surface spin-polarization, reaching values as high as 40% in our samples. We associate the non-trivial magnetic profile to the progressive change of the electronic band structure from a tensile-strained structure at the electrode to a compressive-strained one. The c/a lattice ratio is expected to vary from ~0.985 to ~1.015 across the interfaces passing through $c/a \cong 1$ at an intermediate $t$ value, which is the optimal crystalline structure for double-exchange coupled ferromagnets.[24] The spin-polarization gain is still observed at room temperature in the t-LS0.1MO series while the magnetic order has disappeared for t-LMO samples with $t > t_M$ indicating that the ferromagnetic interaction at the barriers is sensibly dependent on the cation doping and lattice distortion. Our results provide a successful strategy to finely tune the spin-polarization at manganites surfaces by the proper choice of an oxide capping layer.

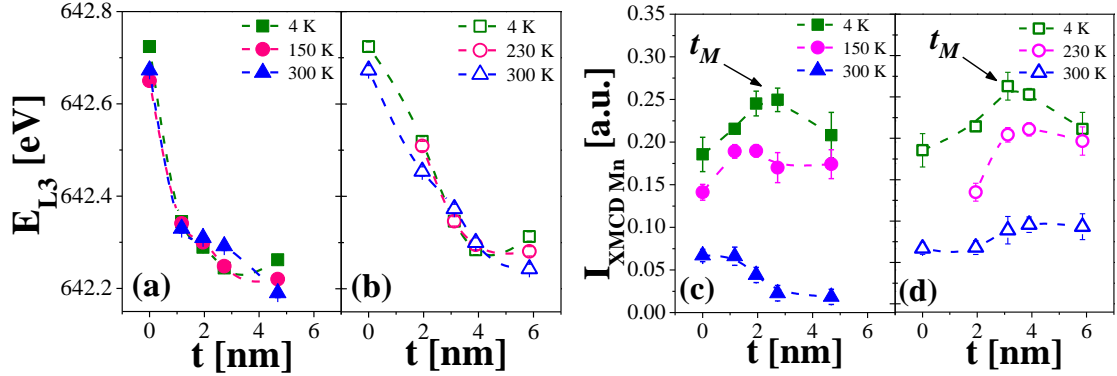

**Figure 2:** $L_3$ absorption energy peak position, $E_{L3}$, as a function of the barrier thickness for (a) t-LMO and (b) t-LS0.1MO samples' series, taken from XAS spectra. Dichroic integrated intensity, $I_{XMCD\,Mn}$ as a function of the barrier thickness for the (c) t-LMO and (d) t-LS0.1MO samples at a temperature range 4-300K.

2.3 Orbital hybridization and La magnetism.

The XAS spectra at the O K-edge probe dipolar transitions from the O 1s to O 2p levels, being the later strongly hybridized with the Mn 3d-states. Thus, the absorption of circular radiation at the O pre-edge region offers valuable information on the density of states and local symmetry of the Mn orbitals. A typical O K-edge XAS spectrum of a LSMO/LSxMO bilayer and that of the LSMO reference sample are shown in Figure 3(a). The first peak of the XAS spectrum (527-533 eV) is characteristic of the O 2p hybridization with the majority $e_g\uparrow$ and minority $t_{2g}\downarrow$ Mn states, whereas the second structure located at ≈531.7 eV corresponds to O 2p-Mn $e_g\downarrow$ states.[25-27] A detailed analysis of pre-edge region (shaded in the spectrum) reveals that the broad peak of the LSMO single layer splits into two well separated ones as the thickness of the capping layer increases. The intensity and energy position of the higher-energy peak increase with the barrier thickness up to 4 nm (≈10 u.c.) when they saturate (Figure 3(b)). Due to the fact that the penetration length is also of the order of 4 nm, we conclude that the saturation values can be attributed to the capping compound.

Although there have been a controversy about the physical origin of the splitting in doped manganites,[28,29] several experiments coincide in assigning the splitting behavior to the combined effects of the crystal field (CF) and Jahn-Teller distortion (JTD).[30,31] However, and in spite of the fact that the JTD plays an important role in the definition of the energy levels structure, the structural distortions induced by strains have a stronger influence on it as shown in Figure 3(b).

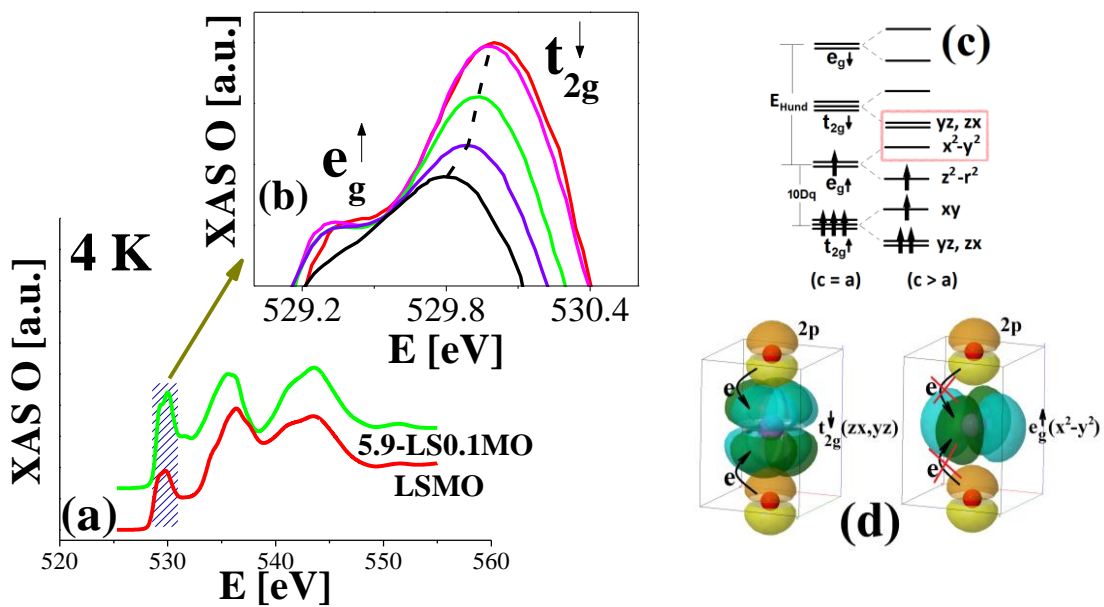

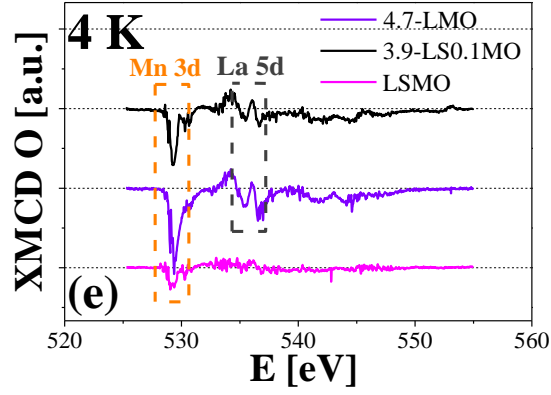

**Figure 3.** (a) XAS spectra measured at the O K-edge in TEY mode for the 5.9-LS0.1MO bilayer and the LSMO reference film. (b) Detail of the XAS at the pre-edge for the whole t-LS0.1MO series of samples at the pre-edge region (denoted in the main panel with a shaded area). (c) Energy diagram for a half filled $Mn^{3+}$ ion, subject to a compressive strain. (d) Schematic of 2p O- 3d Mn orbitals symmetry and hopping. (e) XMCD spectra calculated at the O K edge for the bilayers and the LSMO reference sample. We highlighted the dichroic signal coming from the hybridization of the O K edge with the 3d orbital of the Mn and with the $A^+$ cation.

In particular, the occupancy of the $e_g^\uparrow(z^2-r^2)$ orbitals increases within a compressive-strained structure (c/a > 1) and therefore the empty Mn-3d orbitals available for electronic hopping with O-2p are $e_g^\uparrow(x^2-y^2)$, $t_{2g}^\downarrow(yz)$ and $t_{2g}^\downarrow(zx)$, as is schematically depicted in the energy level configuration shown in Figure 3(c). The orbital hybridization of the $t_{2g}^\downarrow(yz)$ and $t_{2g}^\downarrow(zx)$ states with the out-of-plane O-2p orbitals is more likely due to spatial symmetry arguments (Figure 3(d)), increasing the spectral weight of this configuration at 530 eV, as observed in the experiments. The strong hybridization of Mn 3d $t_{2g}\downarrow$ - O 2p orbitals contributes to diminish the overall Mn magnetic moment at the barrier and would also explain the depression of the spin-polarization observed in samples with large barrier thickness.

Keeping the same experimental conditions as used for the XMCD measurements on the Mn edge, we calculated the ferromagnetic dichroic signal at the O K-edge for the two series of samples and plotted typical curves in Figure 3(e). The negative dichroic signal at the pre-edge region is directly associated with the magnetism of the Mn ion, but another unexpected signal is observed close to 536 eV. This dichroism is related to the 5d La orbitals, giving clear evidence of a spin-polarization at the A-site cation. The unexpected magnetic signal could be explained by the existence of a charge migration through the La-O hybridized orbitals that modify the outer electronic configuration of the cation, endowing it an effective magnetic moment. The same features were observed for both series of samples but they are not in the XMCD of the LSMO reference film.

## 3. Conclusions

In this article we demonstrated by using high-resolution transmission electron microscopy and surface-sensitive techniques that the electronic and magnetic properties at manganite-based interfaces are critically affected by strains and charge distribution within a few nm thick region. The study performed on epitaxial bilayers, $La_{0.67}Sr_{0.33}MnO_3/La_{1-x}Sr_xMnO_3$, show that a finite-length interfacial region is formed in the neighborhood of both compounds that exhibit particular electronic and magnetic properties. A fine tune of them is possible by adjusting the doping level and thickness of the overlayers. The variation of the properties throughout the interface may be sharp or smooth depending on the bilayers´ composition. Circular dichroism measurements, performed at the Mn edge, showed that the spin-polarization of the structures is strongly enhanced for a critical barrier thickness determined by the Sr doping of the barrier. The relaxation of the manganite structures across the interface lead to the stabilization of a pseudo-cubic structure (c~a) in an intermediate region of a few unit cells wide where the double-exchange coupling between $Mn^{3+}$-O-$Mn^{4+}$ prevails. The low-doped manganite barriers that form larger interfaces with gradual charge and orbital variations, give place to higher spin-polarized structures that are still ferromagnetic at room temperature.

Absorption measurements recorded at the O K-edge provide additional information about the effects of the lattice distortion on the electronic band reconstruction and magnetism of the interfaces. On one hand, a variation of the $t_{2g}\downarrow$ and $e_g\uparrow$ multiplets splitting is observed as the c/a ratio increases towards a tetragonal symmetry, together with a progressive enhancement of the 3d Mn - 2p O hybridization through $t_{2g}\downarrow(xz)$ and $t_{2g}\downarrow(yz)$ orbitals. The increase of orbital hybridization also extends to the A-site cations were a magnetic polarization was unexpectedly observed, expressing a new ingredient of the magnetic ordering in perovskite thin films.

## 4. Experimental Section

Two series of $La_{0.67}Sr_{0.33}MnO_3/La_{1-x}Sr_xMnO_3$ (x: 0, 0.1) bilayers were grown on (001) $SrTiO_3$ single-crystalline substrates by pulsed laser deposition. The thickness of the LSMO electrode was kept fix at around 22 nm in both series of samples while the capping layers thickness was varied between 1.2 nm $\leq t_{LMO} \leq$ 4.7 nm in the LSMO/LMO series and 2nm $\leq t_{LS0.1MO} \leq$ 6 nm in the LSMO/LS0.1MO one. In addition, a 20 nm-thick LSMO thin film was grown in similar conditions to use as references. The structural quality of the samples was assessed locally with scanning transmission electron microscopy coupled with a high angle annular dark field detector (STEM-HAADF) in a FEI Titan G2 at 300 keV probe corrected (a CESCOR Cs-probe corrector from CEOS Company) and fitted with a Gatan Energy Filter Tridiem 866 ERS to perform EELS analysis. We used element-specific soft x-ray absorption spectroscopy (XAS) at the $L_{23}$ edge of Mn (630 eV to 670 eV) and at the O K-edge (525 eV to 555 eV) in total electron yield detection mode to probe the magnetism and valence balance throughout the interface between the LSMO electrode and the capping layers. The synchrotron radiation experiments were performed at the electron storage ring of the Helmholtz-Zentrum Berlin (BESSY) by using the 70 kOe high-field end station located at the UE46-PGM1 beamline. The XAS spectra were calculated as the average between the spectra collected with photons of different polarization. The ferromagnetism of the samples was assessed by x-ray magnetic circular dichroism (XMCD) experiments using normal incidence synchrotron radiation with right-handed and left-handed circular polarization. The measurements were performed under a magnetic field of 1T applied perpendicular to the sample surface. The XMCD intensity was calculated as the difference between the absorption of right and left circularly polarized radiation. All the experiments were performed in zero-field cooled samples at three different temperatures: $T_1$ = 4 K, where both the LSMO electrode and the capping layer are ferromagnetically ordered, at $T_2$ = 150 K (230 K) where the LSMO is still ferromagnetic and the LMO (LS0.1MO) is slightly above its Curie temperature and at $T_3$ = 300K, close to the Curie temperature of the electrode.

**Supplementary Material**

In the supplementary material we show the evolution of the compositional profile along the c-axis in the near surface region measured by EELS as well as the Mn $L_{2,3}$-edge absorption spectra measured for the t-LS0.1MO series, where the absorption peaks corresponding to the $Mn^{2+}$ and $Mn^{3+/4+}$ are clearly indicated. Besides, we show the XMCD spectra obtained for the t-LMO series from the right and left circularly polarized radiation.

**Acknowledgements**

The microscopy study have been conducted at the "Laboratorio de Microscopias Avanzadas" from the "Instituto de Nanociencia de Aragón - Universidad de Zaragoza". Authors acknowledge the institution for offering access to their instruments and expertise.

Authors thanks the financial support of FONCYT PICT 2014-1047, CONICET PIP 112-201501-00213, MINCYT and the European Commission through the Horizon H2020 funding by H2020-MSCA-RISE-2016 - Project N° 734187 –SPICOLOST.

**References**

[1] J. Mannhart, and D. Schlom, Science **327**, 1607 (2010).

[2] S. Shi, Z. Sun, A. Bedoya-Pinto, P. Graziosi, X. Li, X. Liu, L. Hueso, V. A. Dediu, Y. Luo, and M. Fahlman, Adv. Funct. Mater. **24**, 4812 (2014).

[3] A. Soumyanarayanan, N. Reyren, A. Fert, and C. Panagopoulos, Nature **539**, 509 (2016).

[4] Y. J. Shin, Y. Kim, S.-J. Kang, H.-H. Nahm, P. Murugavel, J. Rae Kim, M. Rae Cho, L. Wang, S. Mo Yang, J.-G. Yoon, J.-S. Chung, M. Kim, H. Zhou, S. Hyoung Chang, and T. Won Noh, Adv. Mater. **29**, 1602795 (2017).

[5] K.-J. Jin, H.-B. Lu, K. Zhao, C. Ge, M. He, and G.-z. Yang. Adv. Mater. **21**, 4636 (2009).

[6] P. Yu, Y.-H. Chu, and R. Ramesh, Materials Today **15**, 320 (2012).

[7] E. Y. Tsymbal, E. R. A. Dagotto, C. Eom, and R. Ramesh, *Multifunctional Oxide Heterostructures,* 1st ed; Oxford University Press: Oxford, 2012.

[8] S. S. Rao, J. T. Prater, Wu Fan, C. T. Shelton, J.-P. Maria, and J. Narayan, Nano Lett. **13**, 5814 (2013).

[9] M. Bowen, M. Bibes, A. Barthélémy, J.-P. Contour, A. Anane, Y. Lemaître, and A. Fert, Appl. Phys. Lett. **82**, 233 (2003).

**Corresponding Author**

*Laura B. Steren.

Instituto de Nanociencia y Nanotecnología, 1650 Ciudad Autónoma de Buenos Aires, Argentina.

steren@tandar.gov.ar